# Three Dimensional Simulation of Oscillatory Flow in Partial Porous Tube

**Pawan Kumar Pandey**
Department of Mechanical Engineering,
Indian Institute of Technology Kanpur,
Kanpur 208016
Email: pawansut@iitk.ac.in

**Malay Kumar Das**
Department of Mechanical Engineering,
Indian Institute of Technology Kanpur,
Kanpur 208016
Email: mkdas@iitk.ac.in

**Abstract**
Characteristics of Oscillating Flow in a Straight Tube with varying thickness of porous layer at the wall is studied. Results for two different porous layer is discussed and compared in terms of Velocity Profile, phase differences and radial position of maximum velocity over the cross-section. Present problem is modelled using Navier-Stokes Equation in clear medium at the core of tube while Darcy-Brinkmann-Forcheimmer Model is used for the outer porous layer. Solutions for these equations have been obtained using Finite Volume Method based 3D unstructured solver. Velocity profile obtained for tube with high permeability porous layer is the main highlight of results.

## I. INTRODUCTION

Porous layer at the inside wall of a tube appears in variety of flow conditions such as petroleum pipelines and arteries in living beings. Appearance of porous layer may be deliberate and useful or developed with time causing both positive and negative side effects. As in blood carrying arteries porous layer may appear in form of plaque deposition or blood clotting on the site of damaged endothelium layer. Pulsatile nature of blood flow in presence of porous medium creates interesting flow structure which may have haematological effects. Fully developed velocity profile for oscillating flow in straight tube is known as Womersley Profile [1]. Guo et al. in [2] have studied the axisymmetric partial porous tube and presented the velocity profile over the cross-section where thickness of porous layer was 20% of tube diameter. According to work in [2] low permeability cases causes fluid to squeeze out of porous layer and velocity profile resembles as of reduced diameter tube. Although they did not show the velocity profile for fully developed oscillatory flow for high permeability porous layer. Practical Application of Partial Porous Tube model also appears in the porous media layer model for roughness of micro conduits. Roughness of conduit may be dealt in terms of porous property parameters as shown in [3].

Presence of porous layer adds up the resistance in form of viscous and inertial resistance from porous matrix. This is why fully developed velocity profile in Partial Porous Tube will change with the change in porous media layer properties and layer thickness. We could not find any analytical solution of fully developed profile for oscillating flow for general porous and clear media combination in conduit. We were also not able to find the exact expression for flow development length in conduit with porous media layer and clear media combination. Although some expressions are available in [4] for very primitive cases restricted for a narrow range of shape factors.

## II. METHODOLOGY

A Finite Volume Method based solver is used to study the oscillatory flow in partial porous tube. Solver utilizes the unstructured tetrahedral element based mesh for partial porous tube. Flow inside the core of tube is modelled using the Navier-Stokes equation while flow in the outer porous layer is modelled using volume averaged momentum equation for porous media, Darcy-Brinkmann-Forcheimmer equation is adopted from [5]. Fluid rheology is taken as Newtonian. Study in [4] suggest that for Newtonian Fluids viscosity in porous media can be safely assumed same as in clear media.

$$\nabla . \langle V \rangle = 0 \quad (1)$$

$$\frac{\rho_f}{\varepsilon}\left[\frac{\partial \langle V \rangle}{\partial t} + \langle (V.\nabla)V \rangle\right] = -\nabla \langle P \rangle^f + \frac{\mu}{\varepsilon}\nabla^2 \langle V \rangle - \frac{\mu}{K}\langle V \rangle$$
$$-\frac{\rho_f F \varepsilon}{K^{0.5}}[\langle V \rangle.\langle V \rangle]\boldsymbol{J} \quad (2)$$

$$U(t) = U_{osc,max}\sin(\omega t) \quad (3)$$

Where $\langle V \rangle$ the volume average of velocity, F is the inertia co-efficient, $\varepsilon$ is volume fraction of void in porous media, $\boldsymbol{J}$ is the unit vector in velocity vector direction. Other variables have usual meaning corresponding to momentum and mass conservation equations. Eqn. (1) and (2) reduces to mass and momentum conservation equations for clear media when porosity value is 1 and Permeability value is very large. Simulations has been done for Re =100, Wo =11 and sinusoidal velocity variation eqn. (3) at inlet. Schematic of Partial Porous tube used is shown in Fig. 1. For Porous media two sets of properties are used. For high permeability porous media layer $\varepsilon = 0.85$, $K = 10^{-6}\ m^2$ and inertia coefficient =0.25 while for low permeability porous media layer $\varepsilon = 0.6$, $K = 10^{-8}\ m^2$ and inertia coefficient =0.5 is used. Simulation were done at least for total 15 cycles and 3 cycles after dynamic convergence was achieved. Time step size used is $10^{-4}$ seconds and total number of tetrahedral elements used in mesh is around 7,50,000.

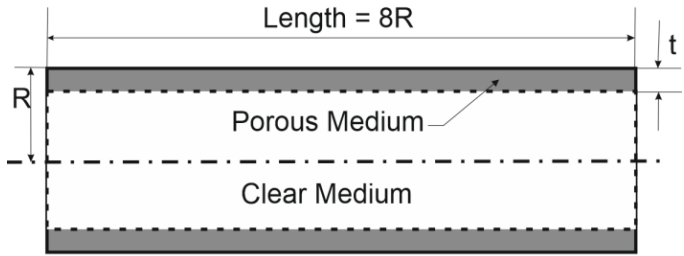

**Fig.1: Diagram of Partial Porous Tube. Shaded region indicates the porous layer at the wall of tube. For 40% Partial Porous Tube t=0.4R.**

### III. RESULTS AND DISCUSSION

Jen et al. [10] have studied the steady flow development in partial porous square channel. We tried to replicate their results for validation of our code. Fig. 2 shows the results of this exercise along with results of Jen et al. [10]. We simulated oscillatory flow for straight circular tube with different thickness of porous layer on the tube wall. To characterise the effect of porous media we used porous properties such as porosity, permeability and corresponding to high permeability and low permeability cases from Paul et al. [7]. Results of high permeability case for porous layer thickness as 40% and 60% of diameter is shown in Fig. 3 and Fig. 4. Maxima in Womersley profile do not occur on the centreline rather it appears on the region where transition between inertial and viscous regime occurs. This position of maxima can be explained by the balance between effect of wall friction and the phase difference of approximately 90° between velocity and pressure waveforms for Womersley number greater than 10. Similar effect is also observed for low and high permeability partial porous tubes. In case of low permeability partial porous tube flow gets squeezed out in the clear core of the tube, consecutively the position of maxima shifts away from porous layer. Velocity waveforms resembles to the oscillatory fully developed profile of diameter reduced by same percentage as of porous layer thickness.

In case of high permeability conditions oscillatory flow fully developed profile in 20% partial porous tube do not exhibit any significant deviation from Womersley profile. Fully developed oscillatory flow profile for 40% and 60% partial porous tube (high permeability) is shown in Fig. 3 and Fig. 4 and they show two step like shape in velocity profile across the cross section. Resistance offered by porous media causes velocity to flatten immediately after the region where wall effect dominates. Inside the core of tube with clear media due to squeezing of fluid from surrounding porous layer again a steep increase in velocity is found. For any instant of time velocity profile have three zones where inertial effects, porous resistance effects and viscous effects come into play with varying degree. After flow waveform attains its maximum, decrease in flow rate starts, pressure starts increasing opposite to the direction of flow. Flow in forward direction and increasing pressure in opposite direction show its maximum effect in porous layer and viscous dominated region near the wall. This effect is evident from flow profile of phase angle $135°$. Fig. 5 shows the difference in the velocity profiles of fully developed flow in completely clear tube, 40% and 60% partial porous tube with high permeability. Fig. 6 shows the variation of phase difference of velocity at any radial position and velocity at the centre of tube. Results are shown for 40% and 60% Partial Porous Tube with high permeability and low permeability conditions for each. Phase difference in velocity for low permeability case becomes very small in porous layer due to high squeezing effect there is a very steep change in phase difference in clear media core. Due to high phase difference opposite flow streams in core and porous layer will occur for longer part of the flow cycle.

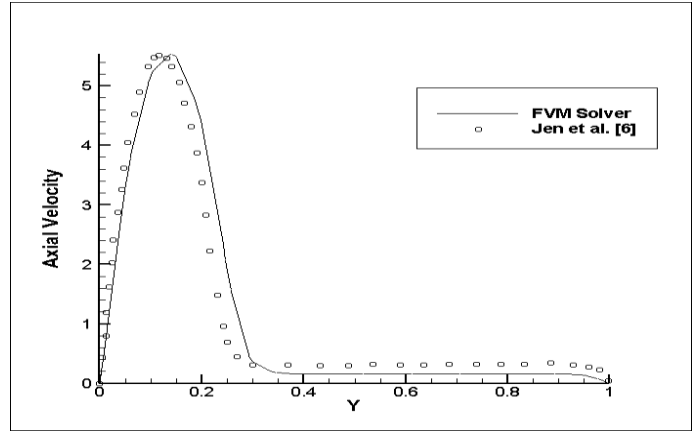

**Fig.2: Comparison of velocity profile obtained for flow in square channel with clear (0<Y<0.25) and porous media (0.25<Y<1.00).**

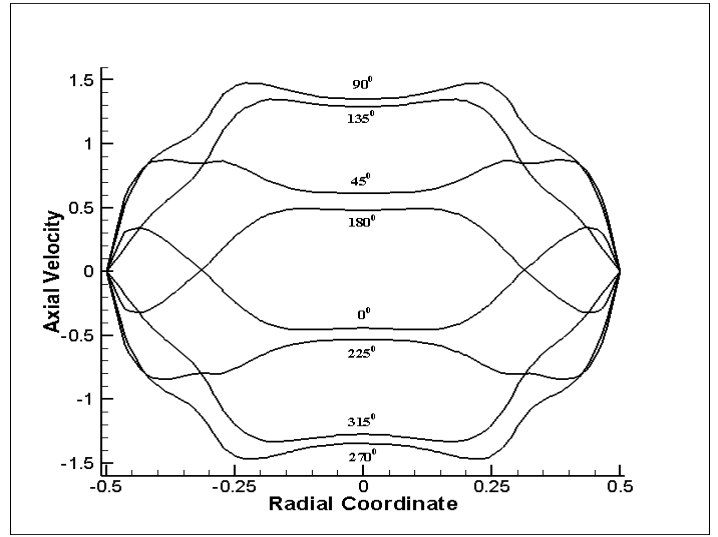

**Fig.3: Axial Velocity profile for oscillatory flow in 40% partial porous tube. Profiles are shown for eight different phase angle at a interval of 45°**

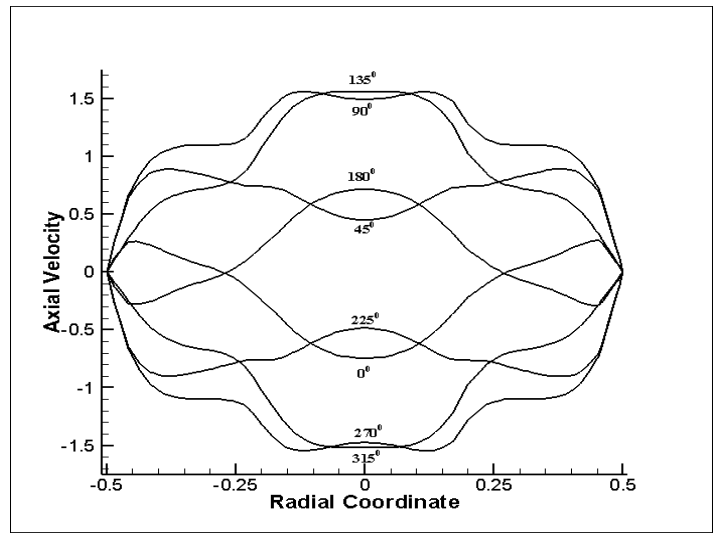

**Fig.4: Axial Velocity profile for oscillatory flow in 60% partial porous tube. Profiles are shown for eight different phase angle at a interval of 45°.**

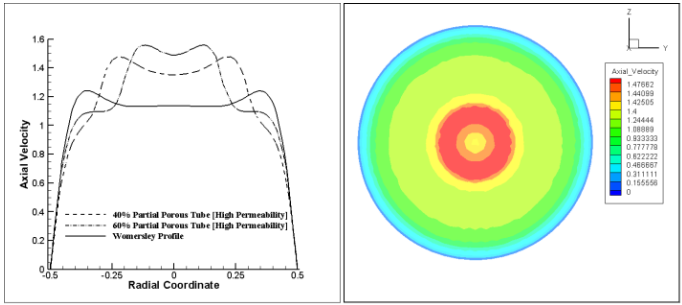

**Fig.5:** Comparison of Axial velocity profile [phase angle: 90⁰] of 40% and 60% Partial Porous Tube with velocity profile for complete clear tube i.e. Womersley profile. Axial Velocity contour for 60% partial porous tube is shown in the right.

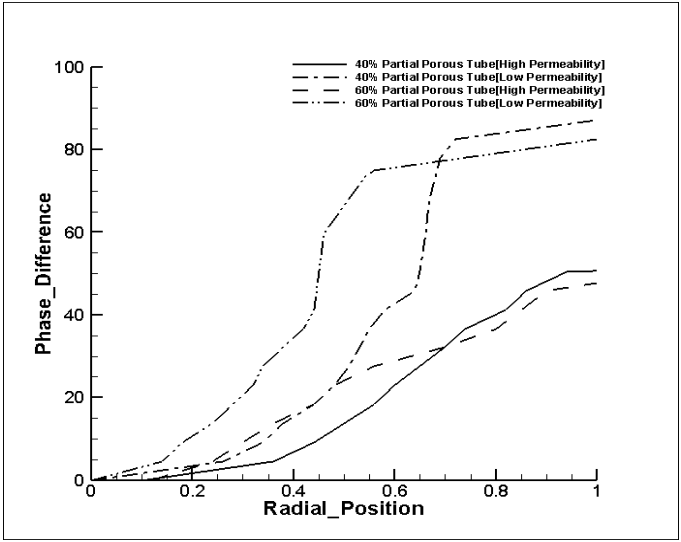

**Fig.6:** Phase difference for velocity along the radius.

## IV. CONCLUSIONS

Fully developed profile for oscillatory flow in Partial porous tube (with porous layer thickness $\geq 40\%$ of diameter) and with high permeability ($> 10^{-6}\ m^2$) porous media shows significant deviations from Womersley profile for clear tube. Squeezing of fluid out of porous layer causes the maxima of fully developed oscillatory profile to shift towards centre as porous layer thickness is increased. Increase in porous layer thickness by 20% of diameter causes around $5°$ reduction in phase difference between velocity at inertia dominated core and friction and wall effect dominated region near the wall. As permeability of porous layer decreases more squeezing effect was observed leading to increased inertial effect in the clear media region causing steep increase in phase differences of velocity at different radial position.

## Acknowledgements

We gratefully acknowledge the High Performance Computing Facility provided by Indian Institute of Technology Kanpur.